\documentclass[12pt]{article}
\usepackage{graphics}
\newcommand{\beq}{\begin{equation}}
\newcommand{\eeq}{\end{equation}}
\newcommand{\pd}{\partial}

\newcommand{\beqs}{\begin{eqnarray}}
\newcommand{\eeqs}{\end{eqnarray}}
\newcommand{\half}{\frac{1}{2}}
\newcommand{\chose}[2]{\left( \begin{array}{c} #1 \\ #2 \end{array} 
\right)}
\newcommand{\symmat}[4]{\left[ \begin{array}{cc} #1 & #2 \\ #3 & #4 \end{array} \right]}
\newcommand{\symmatt}{\left[ \begin{array}{cc} A_{n\times n} & B_{n\times n} \\ C_{n\times n} & D_{n \times n} \end{array} \right]}
\newcommand{\canmat}{\left[ \begin{array}{cc} 0_{n\times n} & I_{n\times n} \\ -I_{n\times n} & 0_{n \times n} \end{array} \right]}
\newcommand{\qh}{\hat{Q}}
\newcommand{\ket}[1]{| #1 \rangle}
\newcommand{\bra}[1]{\langle #1 |}
\newcommand{\braket}[2]{\langle #1 | #2 \rangle }
\newcommand{\poisn}[2]{\{ #1 , #2 \}}
\begin{document}
\bibliographystyle{h-physrev}

\title{Notes on Collective Field Theory of Large N Vector Models as Classical Mechanics on the Siegel Disc\thanks{Based on talk presented  at MRST 2003\cite{mrst2003}}}
\author{A.Agarwal\thanks{abhishek@pas.rochester.edu} \\
University of Rochester. Dept of Physics and Astronomy. \\
Rochester. NY - 14627\\
L.Akant\thanks{lakant@boun.edu.tr}\\
Yeditepe University. Physics Department \\
81120 Kayisdagi - Istanbul}
\maketitle

\begin{abstract}
We use deformation quantization to construct the large N limits of Bosonic vector models as classical dynamical systems on the Siegel disc and study the relation of this formulation to standard results of collective field theory. Special emphasis is paid to relating the collective potential of the  large N theory to a particular cocycle  of the symplectic group. 
\end{abstract}
\section{Introduction:}The purpose of this note is to interpret the large N limit of bosonic vector models as classical dynamics on the Siegel disc $\Sigma (n) = Sp(2n)/U(n)$. We pay special attention to the relation of the collective potentials of the limiting Hamiltonians to one-cocycles of the symplectic group, thus providing a qualitative description of the 'entropy' introduced in the large N theory upon integrating out the non-singlet degrees of freedom. 

No reiteration about the richness of the physical phenomena contained in theories with vector or matrix valued degrees of freedom is really necessary. A key idea in the study of such theories is that of the large N limit\cite{thooft1}. The basic idea behind of the large N limit is the observation that in addition to $\hbar $, which is the measure of the quantum fluctuations of all the observables, there is yet another deformation parameter in the theory, which is $\frac{1}{N}$, which measures only the size of quantum fluctuation of 'gauge' invariant observables. \footnote{These gauge invariant observables, would simply be $O(N)$ invariant bilinears in the case of vector models.} Hence, the large N limit, quite like the $\hbar \rightarrow 0 $ limit should be viewed as a classical limit, albeit an unusual one. Moreover, as  $\hbar $ is usually held fixed in the large N limit, this classical limit is expected to capture various non-perturbative 'quantum' effects, which are missed by the original classical theory that one started with. In other words the large N classical theory is expected to be a better approximation to the full quantum theory than its naive classical limit. To substantiate this idea, it is important to know exactly how non-perturbative quantum information is encoded in the large N limit. In this paper we shall take a small step in this direction.

One of the most successful realizations of a self contained formulation of the large N limit is in the context of matrix models. So before going into the details of vector models, it is worth noting the lessons that we learn there. In the case of the quantum mechanics of a single Hermitian matrix $M$, with the Hamiltonian,
\beq
H = -\hbar ^2\frac{1}{2}\frac{\partial ^2}{\partial M^a_b \partial M^b_a} + Tr V(M),      
\eeq
what one does is to change variables from the matrix elements $M^a_b$ to density of eigenvalues $\rho (x _i)$. The $x_i $'s being the N eigenvalues of the $N\times N$ matrix M. The point of doing this is to have a formulation of the problem in terms of the 'gauge ' invariant quantities, which are the eigenvalues. As was done by Jevicki and Sakita\cite{Collective3, Collective4, Collective5}, one can take the large N limit of the problem, in which the eigenvalues are labeled by a continuous variable $x$ and the resulting Hamiltonian is,
\beqs
H_{N \rightarrow \infty } =  K + V_{coll} + \int dx V(x)\rho (x), \mbox{where,} \nonumber \\
K = \frac{1}{2} \hbar ^2 \int dx \pi'(x)\rho (x) \pi'(x),  \pi(x) = \frac{1}{iN} \frac{\partial }{\partial \rho (x)}, \mbox{and}  \nonumber \\
V_{coll}=\hbar ^2 \frac{\pi ^2}{6} \int dx \rho (x) ^3.
\eeqs
By looking at the limiting large N Hamiltonian, we note that it has the usual kinetic and potential energy terms ($K$ and $V$ respectively), these terms continue to be present when on considers the naive classical limit $\hbar \rightarrow 0$ as well. However, we see that there is an added term; $V_{coll}$ or the collective potential, which  resulted from changing variables from M to $\rho $. The origin of this term lies in the Jacobian of the change of variables. The collective potential is in a sense what makes the large N limit non-trivial. For example if one were to find the ground state of the theory, then,  in the absence of the collective potential one would simply have to solve $V'(x) = 0$, which is  what one would get in the usual classical limit of the theory. The presence of the collective potential  introduces the possibility of ground states which have no analogues in the naive classical limit, but are very special to the large N limit.  In a sense, which was made precise in \cite{abhi-collective, entropy}, the collective potential measures the information that is lost upon integrating out the non-gauge invariant or non-singlet degrees of freedom of the theory. In fact it was interpreted as a natural object in non-commutative probability theory, and was shown to be the 'free Fisher information' discovered by Voiculescu\cite{abhi-collective}. The Jacobian of the change of variable itself has the interpretation of the entropy induced in the problem upon ignoring the non-singlet observables, and it is nothing but the Vandermonde determinant. In terms of $\rho $, this entropy is 
\beq
S(\rho ) = \int dx dy \ln|x-y|\rho (x) \rho (y), 
\eeq
and the collective potential is related to the entropy as 
\beq
V_{coll}(\rho ) = \int dx (\partial _x \frac{\delta }{\delta \rho (x)}s(\rho ))^2 \rho(x) .
\eeq
These identifications helped in extending the collective field theory formalism to the case of the quantum mechanics of several matrices, and the probability theoretic interpretation of the collective potential was helpful in establishing certain lower bounds for the ground states of such multi-matrix models\cite{abhi-collective}. 

In this paper we shall try and illustrate a similar alternative interpretation for the collective potential for vector models. One of the motivations behind doing this is that while dealing with vector models one can directly use field theoretical instead of quantum mechanical models as a starting point.  
We shall be able to see that the phase space of the classical large N system admits global coordinates, and we shall be able to describe it as a quotient of the symplectic group by a unitary group. This quotient space, which is known as the Siegel disc, can be coordinatized by two real symmetric matrices, and we shall be able to use deformation theoretic ideas to derive the canonical Poisson brackets on the Siegel disc. The collective potential, it will turn out, is related to a certain cocycle of the symplectic group, and deriving the precise relation between the cocycle and the collective potential  enables us to have a precise formulation of the entropy introduced in the large N formalism upon integrating out the non-singlet degrees of freedom. 

Various  pieces of the results that we are going to describe in this article have been worked out in the past in many different contexts. For example, the identification of the large N phase space as the Siegel disc has been discussed in several places, in the published literature, see for example \cite{berezin1, berezin-3, rajeev-turgut-qcd}. The classical mechanical formulation of the Large N limit of vector models with either Bosonic or Fermionic degrees of freedom has been studied within the framework of geometric quantization in these and other related papers. The collective field theoretic formalism of Jevicki and Sakita has also been utilized quite extensively in several places \cite{Collective3, Collective4, Collective5, rodrigues-et-al-1}, and a slightly different but equivalent formulation of the problem, using coherent states,  has been discussed by Yaffe \cite{Yaffe1}, and a  recent  thorough review of the large N physics of vector models can be found in \cite{vec-rev}.Part of what we shall do in the paper is to present a unified description of the problem using the language of deformation quantization. Another  observation to keep in mind is that  
the collective potential and its relation to entropy is a feature that does not really depend on the details of the Hamiltonian being studied. It's origin lies in taking the  Large N limit, hence it is a kind of universal object. So it is useful to have a model independent or description of it, which is what we present here. 

From the point of view of the holographic correspondence\cite{adscft}, large N limits of both vector and matrix models are extremely important examples, as these provide rather tractable illustrations of how theories of gravity might emerge from field theories. This is extremely transparent in the case of the quantum mechanical (C=1) matrix model mentioned above, and its string field theory interpretation has been worked out by Das and Jevicki \cite{Stringf1}. Holography makes its appearance in the guise of an extra dimension, which is the dimension of the space of eigenvalues. Large N vector models at criticality too have a conjectured gravity dual, except in this case the dual is not a string theory but rather a higher spin gauge theory of gravity \cite{polyakov-vec}. Recently Das and Jevicki have explored this connection in some detail\cite{das-vec}, see also \cite{mikhailov-hs}. For an extension of this idea in the supersymmetric context, see \cite{nemani-sysy-vec}. The extra dimension appears in this context in the expansion of the bilocal collective fields around the relative coordinate. Moreover, the higher spin interactions manifest themselves in the expansion of the Jacobian of the change of variables in  the relative coordinate as well\cite{das-vec}. Hence, we hope that the geometric interpretation of the Jacobian which we shall present in this paper might be of some use in understanding this conjectured duality better.

\section{Large N Classical Mechanics:} 
Let us outline the basic strategy that will be employed to derive set up the large N limit as a classical limit. The ideas that we shall use basically a translation of those of Yaffe, Berezin and Rajeev \cite{Yaffe1, berezin1, rajeev-neo-classical, govind-thesis} to a deformation theoretic language. We shall start with the quantum mechanical Hamiltonian, $H$ and the associative algebra of observables $\hat {O}^i$'s, generated by some basic canonical commutation relations. Not all the observables will be relevant in the large N limit, and we shall consider only the ones that have leading matrix elements of O(1). The associative algebra can be recast as a deformed algebra with a star product,
\beq
\hat {O}^i \hat {O}^j \equiv O^i \star O^j,
\eeq
where $O^i$'s will be ordinary functions. We shall then compute the leading piece of  algebra of the dominant observables, which in the case of interest will turn out to be a Lie algebra. 
\beq
[O^i , O^j]_{\frac{1}{N} = 0} = \lim_{N \rightarrow \infty}  (O^i \star O^j -  O^j \star O^i)\label{abslie}
\eeq
We shall then identify the coadjoint orbit of this Lie algebra as the limiting classical mechanical phase space. To have an explicit coordinatizatoin of the phase space, 
we shall  proceed to pick a certain set of states $|Z>$, labeled by certain functions  which we shall collectively refer to as $Z$, on which the Lie group acts transitively. The set of functions $Z$'s will then serve as the coordinates on the classical phase space. Given a quantum mechanical operator, $\hat{O}^i$ we shall then be able to associate to it a classical function on the phase space, which is simply the matrix element of $\hat{O}^i$; i.e.
\beq
\hat{O}^i \rightarrow O^i(Z) = <Z|\hat{O}|Z>.
\eeq   
The Poisson brackets between classical functions can be derived as,
\beq
\{O^i(Z), O^j(Z)\} = \lim_{N \rightarrow \infty} (O^i(Z) \star O^j(Z) -  O^j(Z) \star O^i(Z)).
\eeq
This is the abstract of the idea that we shall realize in the context of vector models. It should be noted that the basic idea in this form works for matrix models as well, for example for a recent application of this idea in the context of the AdS-CFT correspondence, we shall refer the reader to \cite{abhi-N=4-1, Abhi-yangian}. 

Let us now focus on the models of interest to us, which are defined by Hamiltonians of the kind,
\beq
\hat{H}/N = \int d^Dx\left(\frac{1}{2}\hat{p}(x).\hat{p}(x) -\frac{1}{2}\partial _x \hat{q}(x).\partial _x \hat{q}(x) + \frac{1}{2}\mu ^2\hat{q}(x). \hat{q}(x) + V(q.q)\right)
\eeq

We shall group together the  position and momentum operators, $q^i(x)$ and $p^i(x)$ $(i = 1 \cdots N)$,   into a single column vector,
\beq
\hat{Q}^i(x) = \chose{\hat{q}^i(x)}{\hat{p}^i(x)}.
\eeq 
For much of the discussion we shall tacitly assume that the spacetime has only a discrete number of points ($n$), and take the continuum limit only at the end.

The observables which dominate in the large N limit are bilinears $X(M)$in $Q$.
\beq
X(M) = \frac{1}{2N} \qh ^i (x) M(x x') \qh ^i(x').
\eeq
Every such $O(N)$ invariant bilinear  is characterized by a symmetric $n \times n$ matrix $M$. It is worth recalling at this point the general fact that the space of real symmetric matrices is isomorphic to the Lie algebra of the symplectic group $sp(2n,R)$.      
\beq
M: M^T = M \Rightarrow J = -\beta M \in sp(2n,R), \beta = \canmat.
\eeq
The $J$'s are elements of the symplectic Lie algebra, as they can be seen to satisfy  $ (\beta J )^T = (\beta J)$. Hence the  invariant observables of the theory are characterized by the symplectic Lie algebra \footnote{As a reference for various properties of the symplectic group and its relation to the Siegel disc, we found \cite{mukunda-et-al-1} to be extremely useful}.  

The bilinears $X$'s  provide a unitary realization of $sp(2n,R)$.
\beq
-i[X(J), X(J')] = X([J,J']).
\eeq
Hence the analog of $\ref{abslie}$ in the context of bosonic vector models is simply the symplectic Lie algebra. 
Upon exponentiation, these bilinears become the generators of linear canonical transformations. Indeed if one defines,
\beq
U(S) = e^{\frac{1}{2}iX(J)}, S = e^J,
\eeq
Then it is relatively straightforward to see that, 
\beq
U^{-1}(S)\hat{Q}U(S) = S\hat{Q}.
\eeq
Keeping in mind that $S$ is an element of the symplectic group and satisfies,
\beq
S\beta S^T = \beta, 
\eeq
one can see that these unitary transformations preserve the canonical commutation relations.

To get the phase space of the classical theory, we shall have to look at the coadjoint orbit of the symplectic group. Explicit coordinates on this coadjoint orbit are provided by  the  Gaussian coherent states. A normalized Gaussian coherent state is parameterized by a symmetric complex matrix $Z: Z^T = Z, Im(Z) >0$. In the coordinate representation,
\beq
\ket{Z} : \langle q \ket{Z} = \left( det U\right)^{1/4}e^{-\frac{1}{2}q(U+iV)q}; Z = V - iU.
\eeq
The space of such Gaussian coherent states is known as the Siegel disc $\Sigma$. The real symmetric matrices $U$ and $V$ are the global coordinates in the disc. Hence the disc is the phase space for the classical theory obtained by taking the large $N$ limit of bosonic vector models. 

The Symplectic group acts on the disc by a generalization of the fractional linear transformations \cite{mukunda-et-al-1, folland}. To be more specific, if 
\beq
S = \symmatt \in Sp,
\eeq
then,
\beqs
U^{-1}(S)\ket{Z} = det^{-1/2}(A -BZ)\ket{Z'}, \\ \nonumber Z' = \alpha (S)Z = (DZ - C)(A - BZ)^{-1} \label{fractional-linear}
\eeqs
The multiplier,
\beq
m(S,Z) = det^{-1/2}(A -BZ),
\eeq
in the equation above is going to be crucial in our later discussions. It has the property of being a co-cycle of the symplectic group. In fact, it may be seen that it satisfies the cocycle condition,
\beq
m(S_1S_2,Z) = m(S_1, \alpha (S_2)Z)m(S_2,Z)
\eeq
%%%%%%%%%%%%%%%%%%%%%%%%%%%%%%%%%%%%%%%%%%%%%%%%%%%%%%%%%%%%%%%%%%%%%%%%%%%%%%%%%%%%%%%%%%%%%%%%%%%%%%
{\bf Constraints on $O(N)$ invariant ovservables:} 

We noted above that the Lie algebra of $O(N)$ invariant observables is the $sp(2n,R)$. As discussed in the preceeding section, that given a real symmetric $2n \times 2n $ matrix $\symmat{a}{b}{b^t}{d}$, we can get an element of $sp(2n,R)$ from it by multiplying it with $-\beta $. Clearly then the dimension of $sp$ is $n(2n+1)$. However, the phase space $\Sigma (n)$ is coordinatized by two real symmetric matrices $U,V$, and hence is $n(n+1)$ dimensional. Indeed, the mismatch has to do with the $n^2$ dimensional subgroup that leave the coherent states unchanged, which is nothing but $U(n)$. Indeed, $\Sigma (n) = Sp(2n)/U(n)$. So, to traverse disc $\Sigma $, by acting on the trivial Gaussian $U =I, V=0$ by exponentials of the bilinears, it is not necessary to consider every possible bilinear. To any bilinear corresponding to an element of $sp$ of the form, $J = \symmat{-b}{-d}{a}{b^t}$, we can add an element of the $U(n)$ Lie algebra, the most general element of which can be written in terms a real skew symmetric $n \times n$ matrix $\lambda $ and a symmetric $n \times n $ matrix $\mu $, $\symmat{\lambda }{-\mu}{\mu }{\lambda }$. Hence, 
\beq
\symmat{-b}{-d}{a}{b^t} \equiv \symmat{-b}{0}{a}{b^t} + \symmat{\lambda }{-\mu}{\mu }{\lambda }
\eeq
Since $a,d$ and $\mu $ are all symmetric, we can chose $\mu $ to cancel out either $a $ or $d$. We shall choose to cancel out $d$. This is a kind of gauge fixing. So it is enough to consider bilinears corresponding to symplectic Lie algebra elements of the type,
\beq
 J = \symmat{-b}{0}{a}{b^t}.
\eeq
This essentially amount to dropping the bilinears which are of the type $p.p$, and working only with those of the kind $q.q$ and $q.p$. 

The corresponding group element is
\beq
S = e^{-\beta J'} = \symmat{\Phi ^{-1}}{0}{\Psi }{\Phi ^T},
\eeq
where,
\beq
\Phi = e^b, \Psi = \int _0 ^1 dt e^{tb^T}ae^{tb}.
\eeq
If the initial state is taken to be the standard Gaussian with $Z = -iI$, then,
using the transformation properties of the coherent states given in (\ref{fractional-linear})
\beqs
V = -\Psi \nonumber\\
U = \Phi ^T \Phi,
\eeqs
and, the cocycle is,
\beq
m(S,Z) = det ^{-1/2}(\Phi ^{-1})= e^{\frac{1}{4}Tr \ln  U}.
\eeq
We shall come back to this formula a little later to relate the collective potential to the cocycle. 

\subsection{ Deformation Theoretic Construction of Observables of the Classical Large N Theory: }

To any observable in the quantum theory ($\hat{A}(\hat{q}, \hat{p}))$, one can associate a function on the disc  $A_c$. The function being simply the coherent state expectation value of the operator in the quantum theory.   
\beq
i.e. \hat{A} \rightarrow A_c = \langle Z|\hat{A}\ket{Z}. 
\eeq
It is useful to use some deformation-theoretic ideas to compute these expectation values. For a recent discussion of the deformation theoretic ideas to the phase space formulation of quantum mechanics see \cite{zachos-qm-phase-space}. Let us recall that  
\beq
\langle Z|\hat{A}\ket{Z} = Tr \hat{A}\rho _Z, 
\eeq
where, $\rho $ is the coherent state density matrix 
\beq
\rho_Z =  \ket{Z}\langle Z|.
\eeq
A result from deformation theory that, for any operator $\hat M$ and $\hat N$, one has,
\beq
Tr\hat{M} = \int d^{2n}Q W_M
\eeq
where, $W_M$ in the Weyl symbol of $\hat{M}$, i.e.
\beq
W_M(Q)= \int d^nq' \langle q -\frac{1}{2} q'|\hat{M}\ket{q + \frac{1}{2}q'}e^{ipq'},
\eeq
Moreover, symbols of products of operators are related to the individual symbols of the operators by the Groenwold-Moyal star product. Or in other words, for any two operators $\hat{M}$ and $\hat{N}$,
\beq
W_{MN} = W_M \star W_N  = W_M(p,q)e^{\frac{i\hbar }{2}(\stackrel{\leftarrow}{\pd _q}\stackrel{\rightarrow}{\pd _p} - \stackrel{\leftarrow}{\pd _p}\stackrel{\rightarrow}{\pd _q})}W_N(p,q).
\eeq
With the above results from deformation theory in mind, the rule of association of a classical function on the disc to an observable of the quantum theory can be sussinctly summarized as,

\beq
\hat{A} \rightarrow A_c = \langle Z|\hat{A}\ket{Z} = Tr \hat{A}\rho _Z = \int d^{2n}Q W_A \star W_{\rho _Z}\label{correspondence}
\eeq
It is to be kept in mind that the star product used above corresponds to symmetric or Weyl ordering. 
The symbol for any polynomial in the position and momentum operators can simply be obtained by re placing the (Weyl ordered )operators in the polynomial by the corresponding commuting quantities i.e.    

\beq
\hat{A}(\hat{p}, \hat{q})_{Weyl-Ordered} \rightarrow W_A = A(q,p),
\eeq
and,
The Weyl symbol for the coherent state density matrix can be computed using the position space representation of the coherent state. A direct computation gives \cite{mukunda-et-al-1},
\beq
W_{\rho _Z} = e^{-QG(U,V)Q}, G = \symmat{U + VU^{-1}V}{VU^{-1}}{U^{-1}V}{U^{-1}}.
\eeq
The function on the disc corresponding to the standard bilinears may be computed using the formalism outlined in (\ref{correspondence}), and the explicit forms of the necessary Weyl symbols given above. The results are,

\beq
\hat{A}(\alpha ,\beta ) = \frac{1}{2}\hat{q}_i(\alpha )\hat{q}_i(\beta ) \rightarrow A_c = \frac{1}{4}U^{-1}(\alpha , \beta );
\eeq   
and 
\beq
\hat{B}(\alpha ,\beta ) = \frac{1}{2}(\hat{q}_i(\alpha )\hat{p}_i(\beta ) + \hat{p}_i(\beta )\hat{q}_i(\alpha ))\rightarrow B_c = -\frac{1}{2}(VU^{-1})(\alpha , \beta ).
\eeq   
\beq
\hat{C}(\alpha ,\beta ) = \frac{1}{2}\hat{p}_i(\alpha )\hat{p}_i(\beta )\rightarrow C_c = \frac{1}{2}\left(VU^{-1}V + \frac{1}{2}U\right)(\alpha , \beta );
\eeq
Clearly, the Hamiltonian of the system the vector models is built out of these bilinears, so the results above are enough to formulate the Hamiltonian as a function on the disc. Or in other words, we now have a deformation theoretic way deriving the classical Hamiltonian of the  limiting large $N$ theory. To get any dynamical information out of this theory, we need to derive the Poisson brackets on the disc.The matrices $U,V$ are the global coordinates on the phase space of the classical theory corresponding to the Large N limit, so we proceed to derive the Poisson brackets between them.

{\bf The Poisson Brackets on the Siegel Disc:}

A reasonable definition of the Poisson brackets between function on the disc is.
\beq
\poisn{A_c}{B_c} \stackrel{def}{=} \int d^{2n}Q\lfloor \left( W_A \star W_B - W_B\star W_A\right)\rfloor \star W_{\rho _Z}, 
\eeq
where the floor implies that only  leading order terms in $1/N$ are kept in the star commutator.   
We now note that
\beq 
\poisn{A_c}{B_c} = -\frac{1}{8}\poisn{U^{-1}}{V}U^{-1}.
\eeq
Comparing this to the result obtained from the direct calculation implied by the definition of the Poisson bracket gives us,
\beq
\poisn{U^{-1}(\alpha, \beta )}{V(\gamma ,\delta )} = -(\delta _{\gamma , \alpha }\delta _{\beta , \delta } + \delta _{\alpha , \delta }\delta _{\beta , \gamma }) \label{poisson}
\eeq 

This completes the abstract program of translating the Large $N$ limit of bosonic vector models as  classical dynamical systems. To every gauge invariant observable of the quantum theory, we now associate a function on the disc as per (\ref{correspondence}), and study the dynamics, of the classical system with the Poisson brackets given by (\ref{poisson}).

{\bf The Hamiltonian:}The Hamiltonian for a bosonic vector model in $D$ space dimensions is,
\beq
\hat{H}/N = \int d^Dx\left(\frac{1}{2}\hat{\pi}(x).\hat{\pi}(x) -\frac{1}{2}\partial _x \hat{\phi}(x).\partial _x \hat{\phi}(x) + \frac{1}{2}\mu ^2\hat{\phi}(x). \hat{\phi}(x) + \frac{\lambda }{4}(\hat{\phi}(x). \hat{\phi}(x))^2\right)
\eeq
Using the results mentioned above, one can now associate to this quantum Hamiltonian, a classical one, which is a function on the disc. The classical Hamiltonian is,
\beq
H_c = \frac{1}{2}Tr\left(\frac{1}{2}VU^{-1}V + \frac{1}{2}U + \frac{1}{2}(-\nabla ^2_x + \mu ^2)U^{-1} + \frac{\lambda }{8} U^{-2}\right),
\eeq
where the trace is to be understood in the sense of matrix multiplication. In the continuum notation used above, one is to understand, for example $TrVU $ as $\int d^Dxd^Dx'V(x,x')U(x',x)$, and $Tr  \nabla ^2 U^{-1}  = \int dx \nabla _x U^{-1}(x,y)|_{x=y}$. This is the standard Hamiltonian that one gets using methods of collective field theory as well. The critical points of the Hamiltonian are of special interest , and one can see right away that the Hamiltonian is extremized at $V =0$, and $U=U_0$, given by the gap equation,
\beq
U_0^{-1}(x,x') = \int \frac{d^Dk}{(2\pi)^D}(k^2 + \mu ^2 + \lambda \sigma )^{-1/2}e^{ik(x-x')},
\eeq
where, the gap, $\sigma (x) = \frac{1}{2}U^{-1}(x,x)$, is to be determined in a self-consistent way
by letting $x \rightarrow x'$ in the equation above. 

The existence of the gap in the spectrum, obtained here as a static solution of the classical theory, is one of the many indications that the limiting classical theory obtained in large $N$ limit a better approximation to the full quantum theory, than the classical theory  that one started with. This non-trivial static solution was possible due to the presence of the term $\frac{1}{4}TrU$, in the Hamiltonian. If we trace back out steps, we can easily see that this term was the result of taking the coherent state matrix element of $p.p$. This term is the contribution of the Jacobian of the change of variables from the position and momentum operators to the $O(N)$ invariant observables, which are the coordinates on the disc. The transformation of the kinetic energy, contributes to the effective potential of the new classical theory, allowing it to generate non-trivial static solutions. In the language of collective field theory, this term is nothing but the collective potential,
\beq
V_{coll}(x) = \frac{1}{4}U(x,x)
\eeq
From our previous analysis, we can relate the collective potential to the cocycle as,
\beq
V_{Coll}(x) = -\frac{\partial}{\partial U^{-1}(x,x)}\ln m(S,Z)
\eeq
To complete the analogy with matrix models mentioned in the introduction, we see that the role of the density of eigenvalues is played, roughly speaking by the bilocal field $U(x,x')$. The cocycle measures the 'entropy' of integrating out the non-singlet degrees of freedom, and the collective potential is related to it through a logarithmic derivative. In the appendix we sketch out the connection of the cocycle to Jacobians. It is also worth noting that although the interpretation of the cocycle as an entropy makes sense from a statistical mechanical point of view, it is not the same notion of entropy that one builds out of counting the states of the field theory. From that point of view, the cocycle is more like a zero point energy. 

%%%%%%%%%%%%%%%%%%%%%%%%%%%%%%%%%%%%%%%%%%%%%%%%%%%%%%%%%%%%%%%%%%%%%%%%%%%%%%%%%%%%%%%%%%%%%%%%%%%

\section{Appendix:}
{\bf Relation of the Cocycle to Jacobians:}

Any real matrix belonging to $Sp(2n,R)$ can be built out of matrices of the type, $\symmat{A}{0}{0}{A^{-1}}$, $\symmat{I}{0}{C}{I}$ and $\symmat{0}{I}{-I}{0}$. This is nothing but a kind of polar decomposition. 
An illustration of the fact that the origin of the cocycle lies in a Jacobian, we shall outline the derivation of the action of the symplectic group on Gaussian coherent states. We shall work out only the case of
\beq
S = \symmat{A}{0}{0}{A^{-1}} \in Sp(2n,r),
\eeq 
in detail, as this is only non-trivial part. The other two types of symplectic matrices simply contribute phases to the cocycle. For a more thorough discussion of these properties we shall refer the reader to \cite{folland}.
We want to prove that 
\beq
U(S)f(x) = \beta \frac{1}{\sqrt{A}}f(A^{-1}x),
\eeq
where $\beta $ is an undetermined complex number of unit modulus. To prove this we shall use the transformation property of Weyl symbols under symplectic transformations, which is,
given an operator
\beq
\hat{A} \rightarrow \hat{A}' = U(S)\hat{A}U^{-1}(S), W_A(Q) \rightarrow W_{A'}(Q) = W_A(S^{-1}Q).
\eeq
Let us now form the density matrix associated with  the vector $\ket{f}$.
\beq
\ket{f} \rightarrow \hat{f} = \ket{f}\bra{f}.
\eeq
The associated Weyl symbol is
\beq
W_f(Q) = \int dx' \braket{x-\half x'}{f}\braket{f}{x + \half x'}e^{ix'p}.
\eeq
Taking into account the special form of the symplectic matrix, it follows that the Weyl symbol transforms into,
\beq
W_f \rightarrow W_{f'} = \int dx' \braket{A^{-1}x-\half x'}{f}\braket{f}{A^{-1}x + \half x'}e^{ix'Ap}
\eeq
We can now change the integration variable form$x'$ to $Ax'$ and absorb the Jacobian of the change of variable $detA ^{-1}$ into a redefinition of the vector/wavefunction.i.e.
\beq 
W_{f'} = dx'\int \frac{1}{\sqrt{A}}f^{*}(A^{-1}(x-\half x')\frac{1}{\sqrt{A}}f(A^{-1}(x+\half x')e^{ix'p}.
\eeq
Which allows us ro read off $f'$ upto a phase,
\beq
f'(x) = \frac{\beta }{\sqrt{A}}f(A^{-1}x), |\beta | = 1.
\eeq
Now taking $|f>$ to be $|Z>$, the result follows. This is the deformation theoretic way of realizing the absorption of the Jacobian in a redefinition of the wave function, which is a familiar tool in collective field theory. 

{\bf Acknowledgments:}  We are indebted to  Sumit Das and S.G.Rajeev for many useful discussions, and for encouraging us to organize our modest results into this paper.  AA is gateful to Sumit Das for his comments and correspondences about an earlier version of this manuscript.  This work was supported in part by US Department of Energy grant number DE-FG02-91ER40685

\bibliography{abhishekbib}
\end{document}